\documentclass{PoS}

\title{Initial state and hydrodynamic modeling of heavy-ion collisions at RHIC BES energies}

\ShortTitle{Dynamical modeling of heavy-ion collisions at RHIC BES energies}

\author{\speaker{Chun Shen}\\
        Physics Department, Brookhaven National Laboratory, Upton, NY 11973, USA\\
        E-mail: \email{cshen@bnl.gov}}

\author{Bj\"orn Schenke\\
        Physics Department, Brookhaven National Laboratory, Upton, NY 11973, USA\\
        E-mail: \email{bschenke@bnl.gov}}

\abstract{We present a fully three-dimensional initial state model for relativistic heavy-ion collisions at RHIC Beam Energy Scan (BES) collision energies. The initial energy and net baryon density profiles are produced based on a classical string deceleration model. The baryon stopping and fluctuations during this early stage of the collision are investigated by studying the net baryon rapidity distribution and longitudinal decorrelation of the transverse geometry.}

\FullConference{Critical Point and Onset of Deconfinement - CPOD2017\\
		7-11 August, 2017\\
		The Wang Center, Stony Brook University, Stony Brook, NY}

\begin{document}

\section{Introduction}

The experimental heavy-ion programs, such as the Relativistic Heavy-Ion Collider (RHIC) Beam Energy Scan (BES) program and the NA61/SHINE experiment at the Super Proton Synchrotron (SPS), provide us with a unique opportunity to explore the phase diagram of strongly interacting matter. A wide range of collision energy ensures extensive coverage in the temperature and baryon chemical potential of the produced nuclear matter. The wealth of hadronic measurements from the RHIC BES \cite{Adamczyk:2017iwn} contain valuable information about the thermodynamic and transport properties of the quantum chromodynamic (QCD) matter in a baryon rich environment. These aspects are challenging to study from first principles lattice computations. The phenomenological study of theory-data comparisons, on the other hand, plays an important role in elucidating the properties of the strongly-coupled Quark-Gluon Plasma (QGP) \cite{Heinz:2013th, Gale:2013da}. Moreover, the search for a critical point in the nuclear matter phase diagram relies crucially on a detailed understanding of the dynamics of collision systems.

The hybrid (viscous hydrodynamics + hadronic transport) model has been shown to be an effective theoretical framework for heavy-ion collisions over a wide range of collision energies \cite{Song:2010mg,Karpenko:2015xea}. Relativistic viscous hydrodynamics describes the macroscopic dynamics of the high density phase of the collision system. It is then switched to a transport description which captures the dynamics of a dilute hadronic phase microscopically.

However, an additional challenge arises at low collision energies, $\sqrt{s_\mathrm{NN}} \sim 10 $ GeV, where the relativistic Lorentz contraction factors of the colliding nuclei along the beam (longitudinal) direction are not large. The finite extensions of the colliding nuclei are illustrated in Fig.~\ref{fig1} for central Au+Au collisions at different collision energies. This finite longitudinal thickness leads to a substantial overlapping time $\tau_\mathrm{overlap} \sim 2-3$ fm/$c$ during which the nucleons inside one nucleus collide with those from the other nucleus. This leaves a large theoretical uncertainty for the early time dynamics of the collision system. Hence, one could expect that the modeling of baryon stopping and density fluctuations along the longitudinal direction during this pre-equilibrium stage has substantial influence on the subsequent hydrodynamics + hadronic transport evolution.
A previous work \cite{Karpenko:2015xea} used the hadron transport model UrQMD \cite{Bass:1998ca,Bleicher:1999xi} to evolve the system during the this pre-equilibrium stage and then connected to hydrodynamics after the overlapping time. A similar approach with the AMPT transport + hydrodynamic simulations was presented for high energy heavy-ion collisions in Refs. \cite{Pang:2012he,Pang:2015zrq}.

Our work \cite{Shen:2017bsr} proposed a new dynamical framework which connects the pre-equilibrium stage of the system and hydrodynamics on a local collision-by-collision basis. The hydrodynamic evolution starts locally at a minimal thermalization time after the first nucleon-nucleon collision. The sequential collisions between nucleons that occur after will contribute as energy and net-baryon density sources to the hydrodynamic simulations. 
In this proceeding, we provide some complementary results to Ref.~\cite{Shen:2017bsr}.

\section{A fully 3D dynamical initialization framework}

At collision energies relevant for the RHIC BES, we need to consider the finite size of the collision zone between the two incoming nuclei. This requires us to set up a framework to model the spatial and momentum configurations of the participating nucleons in 3D event-by-event. In Ref. \cite{Shen:2017bsr}, we generalized the conventional Monte-Carlo Glauber model by considering the spatial distribution of the nucleons along the longitudinal direction. 
\begin{figure}[t]
\begin{center}
 \includegraphics[width=0.95\linewidth]{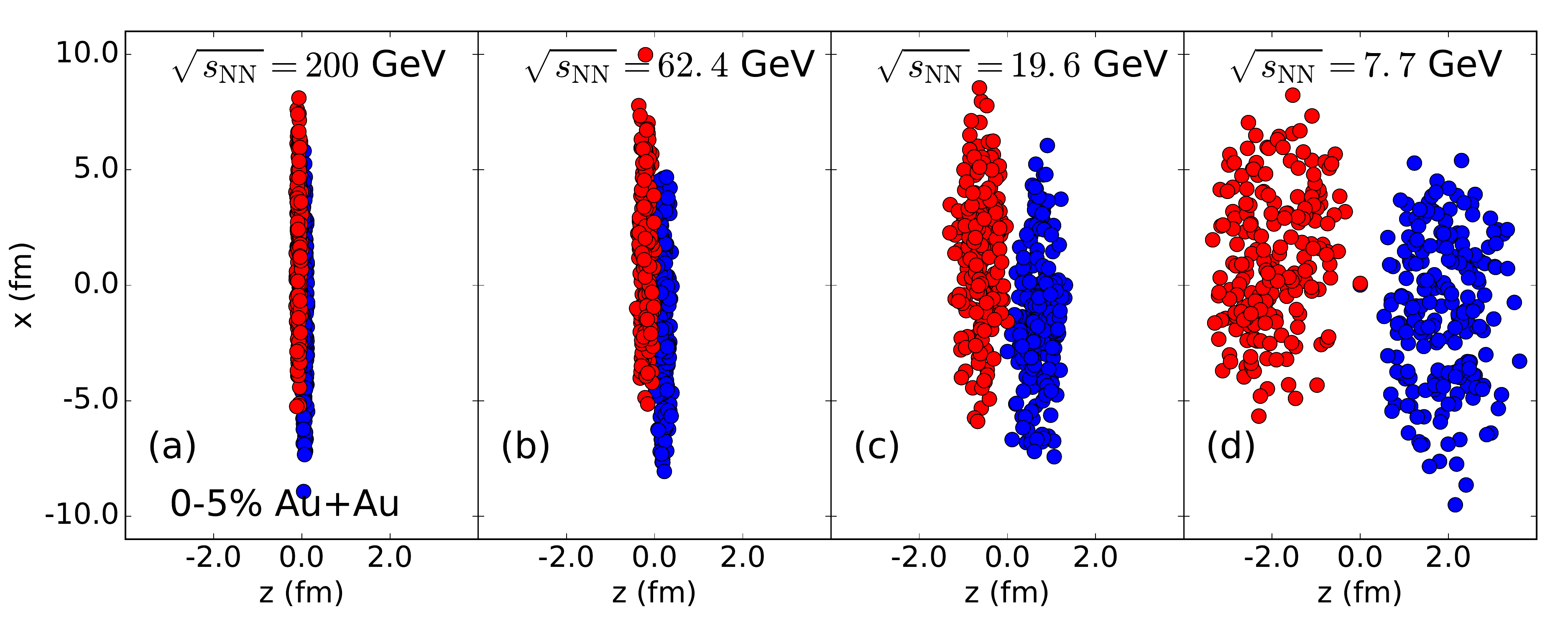}
\caption{The spatial configurations of nucleons inside the colliding nuclei in 0-5\% Au+Au collisions at 4 collision energies. }
\label{fig1}
\end{center}
\end{figure}
Fig.~\ref{fig1} demonstrates the nucleon spatial configurations along the beam direction for Au+Au collisions from 200 GeV to 7.7 GeV. One can clearly see that the Lorentz contraction becomes weaker with decreasing collision energy.
\begin{figure}[b!]
\begin{center}
\begin{tabular}{cc}
 \includegraphics[width=0.45\linewidth]{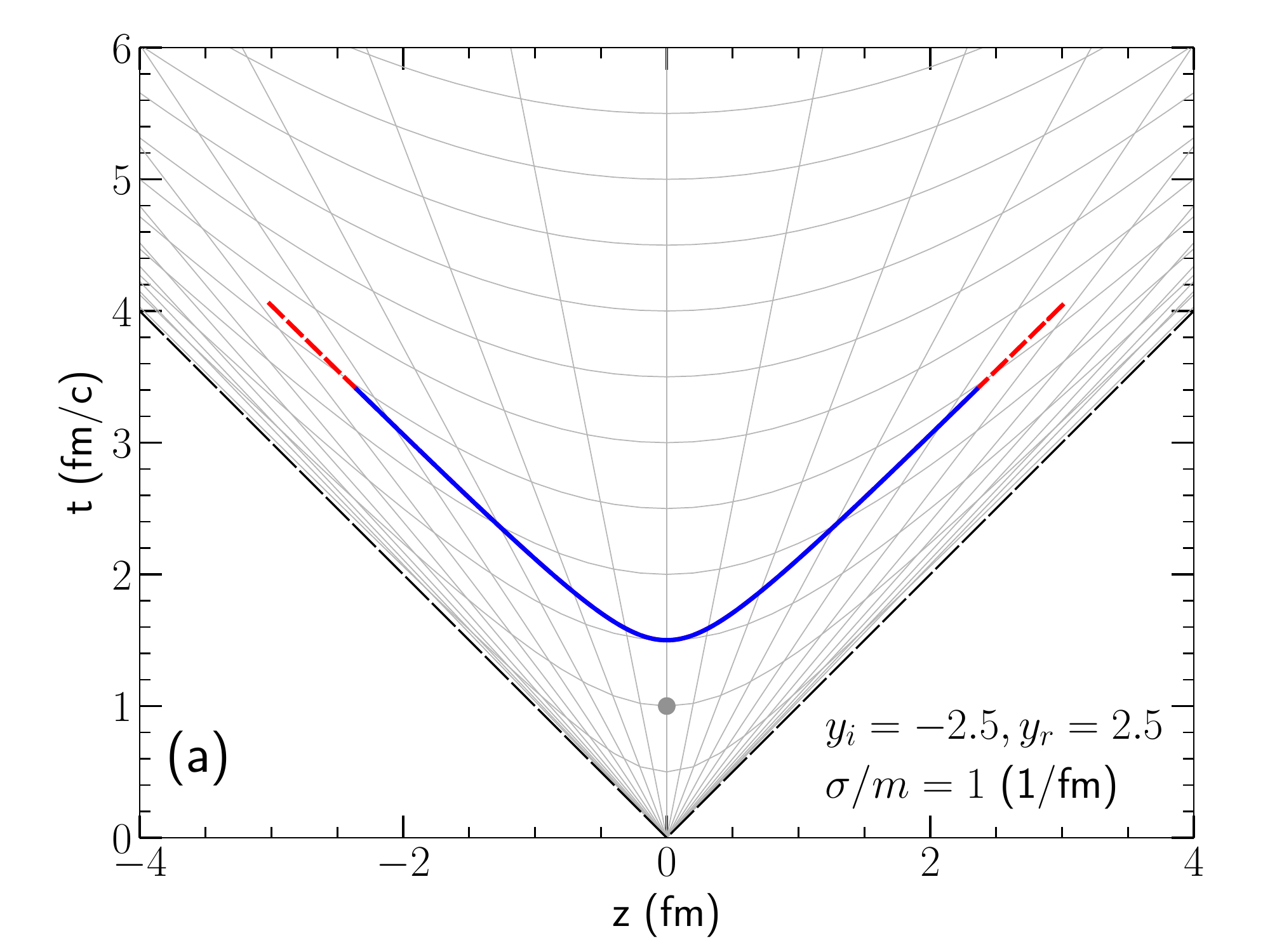} &
 \includegraphics[width=0.45\linewidth]{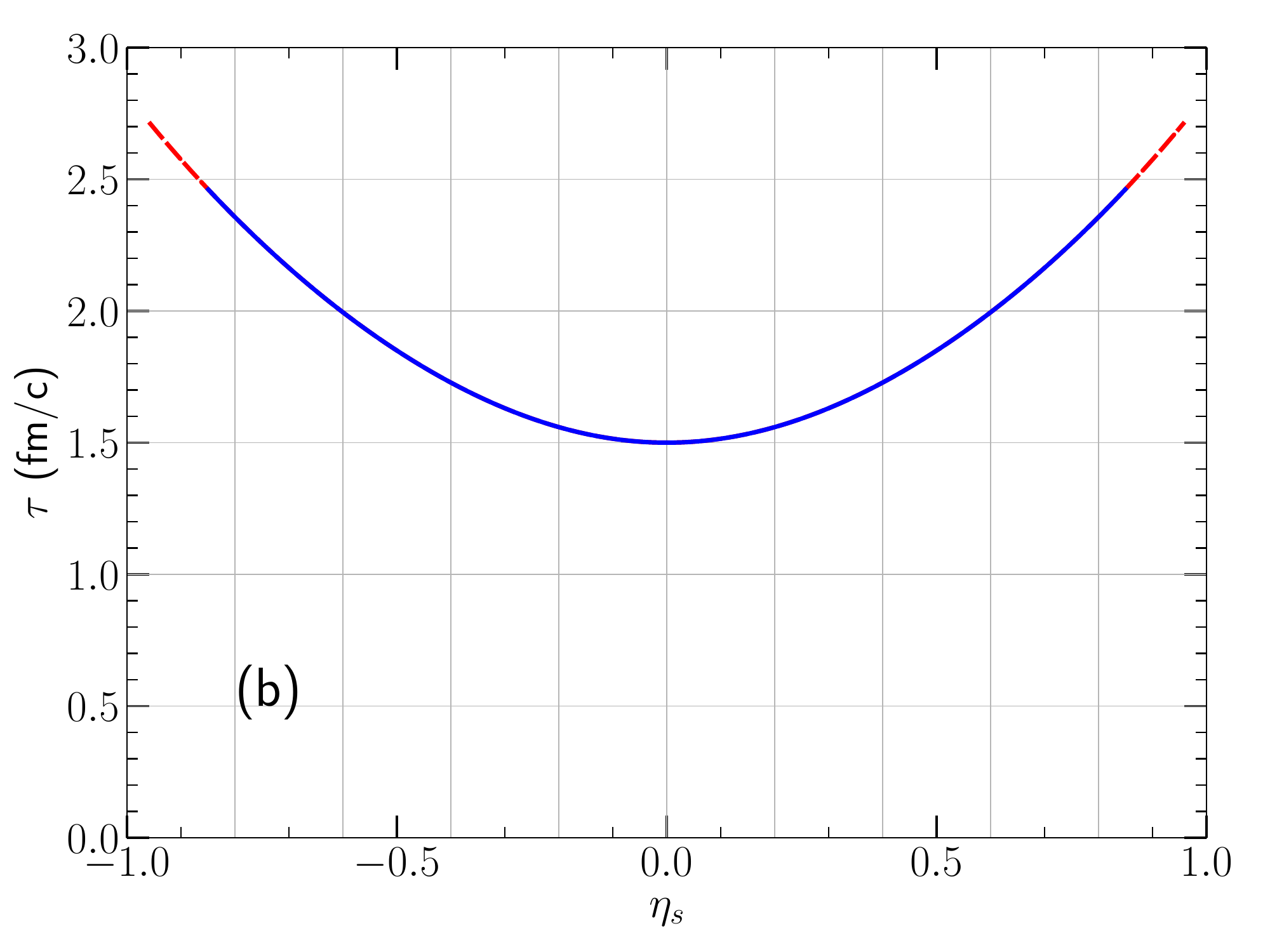}
\end{tabular}
\caption{The space-time distribution of a decelerated string in $t-z$ and $\tau-\eta$ coordinates. The string is produced at $t_0 = 1$ fm/c and $z_0 = 0$ fm indicated by the grey point. The string was evolved for $\Delta \tau = 0.5$ fm/c. The blue line indicates the string position after evolution. The red dashed line represents the case where the string ends were evolved with free-steaming dynamics instead of deceleration. The grey grid in panel (a) shows lines of constant $\tau$ and $\eta$.}
\label{fig2}
\end{center}
\end{figure}
To generate the spatial and momentum distributions for the wounded nucleons after their collisions, we employ the string deceleration model proposed in Ref.~\cite{Bialas:2016epd}.
The collided nucleons form the string end points and are decelerated according to the following classical equations of motion,
\begin{equation}
\frac{dE}{dz} = -\sigma \quad\mbox{ and }\quad \frac{dp_z}{dt} = - \sigma. 
\end{equation}
Here $\sigma$ is the string tension.
Longitudinal fluctuations are introduced by considering the valence quarks, which carry only a fraction of the nucleon's momentum, as the colliding participants and by fluctuating the string deceleration time according to the phenomenologically motivated LEXUS model \cite{Jeon:1997bp}.
The space-time distribution of a decelerated string is shown in Fig.~\ref{fig2}. The string distribution exhibits a convex structure in $\tau-\eta$ coordinates, in which the relativistic hydrodynamic simulation is formulated. This is because the string is produced at $t > 0$. As shown in Fig.~\ref{fig2}a, the string's $t-z$ distribution has a larger curvature compared to the constant $\tau$ hyper-surfaces near it. The large $\vert z \vert$ portion of the string intercepts a larger constant $\tau$ hyper-surface, than the part at smaller $\vert z \vert$.
Compared to a string evolved according to free-streaming, the deceleration dynamics leads to a shorter string at its thermalization. 

\section{Results and discussion}

\begin{figure}[h!]
\begin{center}
\begin{tabular}{cc}
 \includegraphics[width=0.48\linewidth]{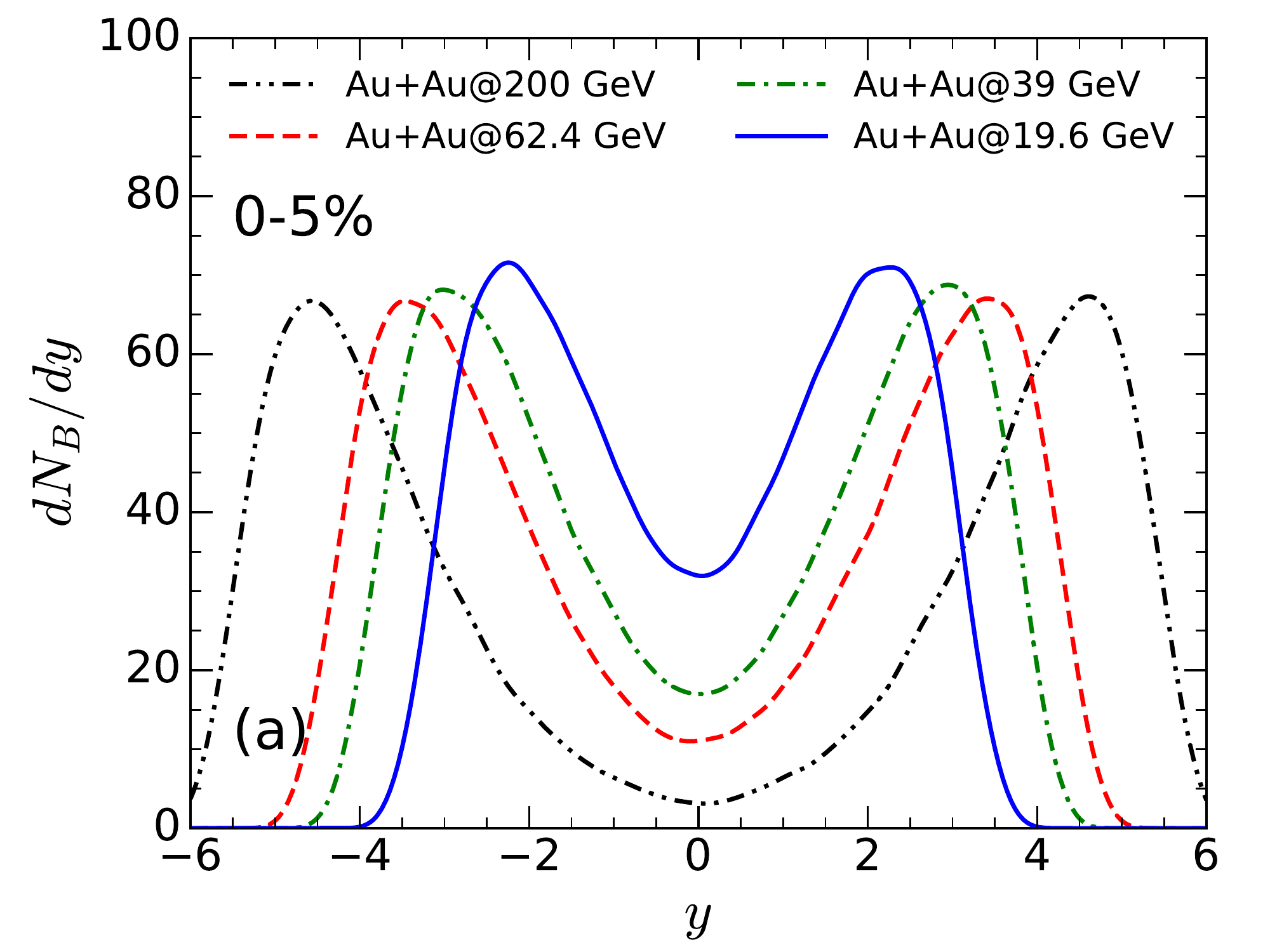} &
 \includegraphics[width=0.48\linewidth]{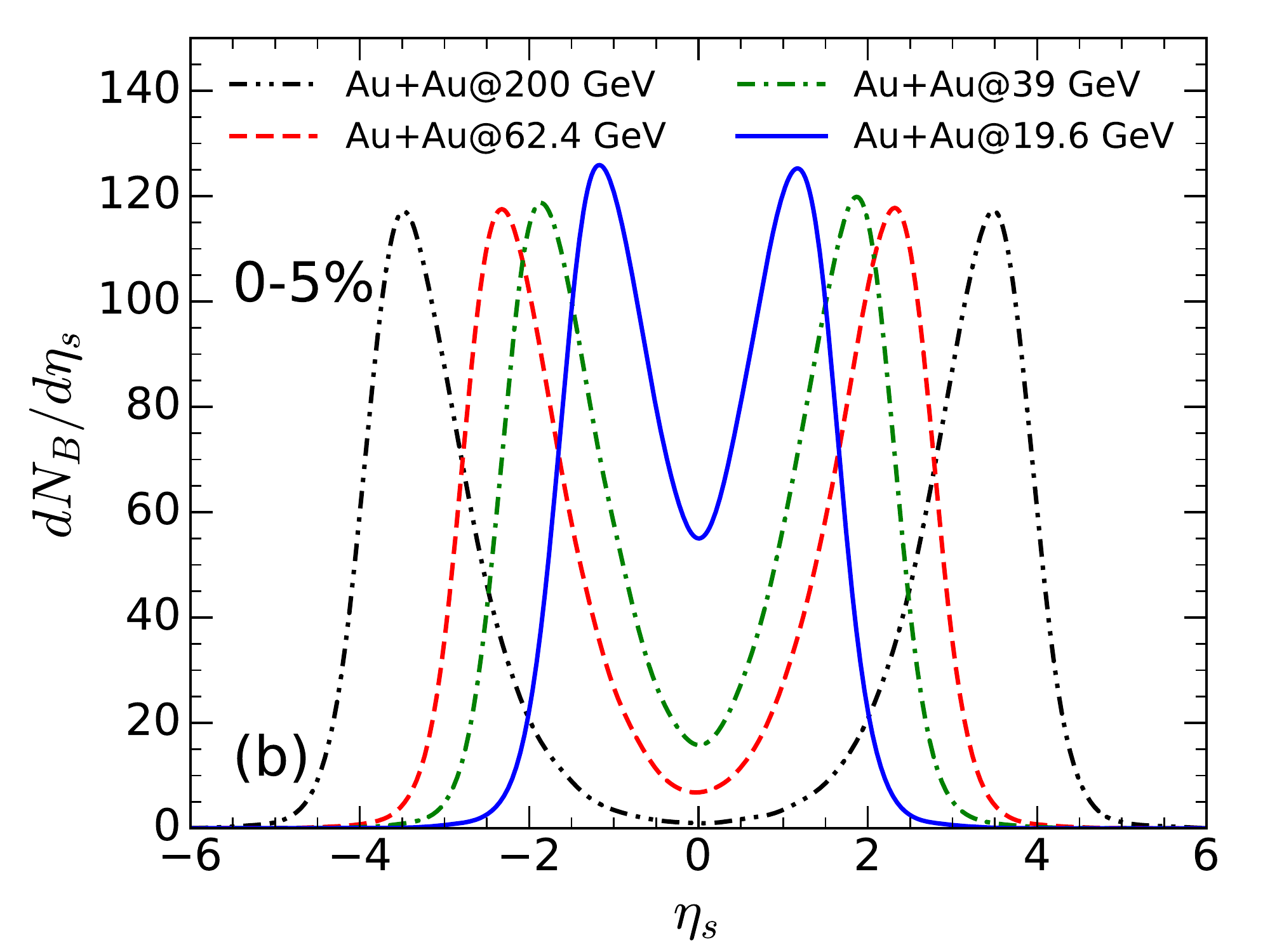} \\
 \includegraphics[width=0.48\linewidth]{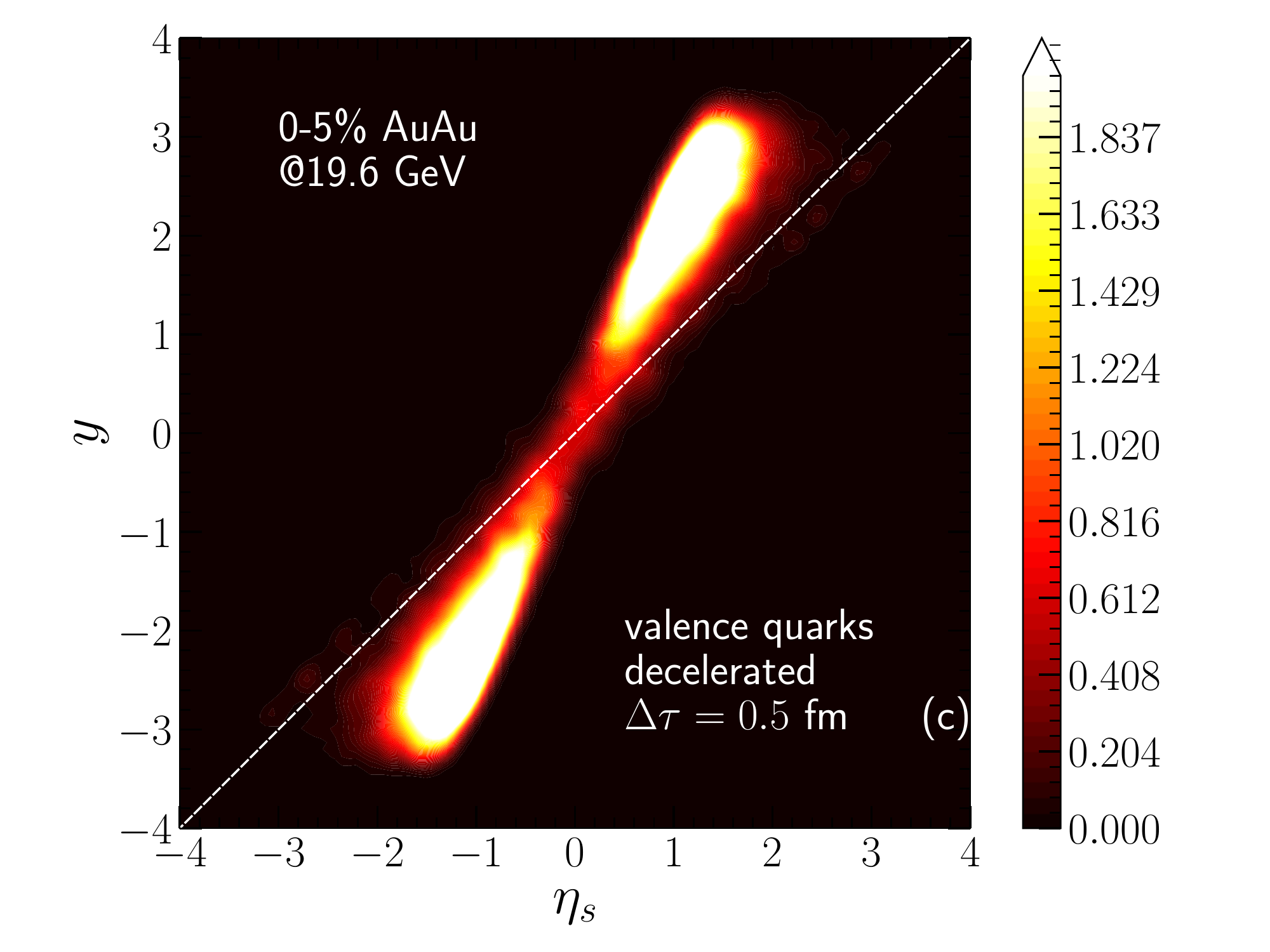} &
 \includegraphics[width=0.48\linewidth]{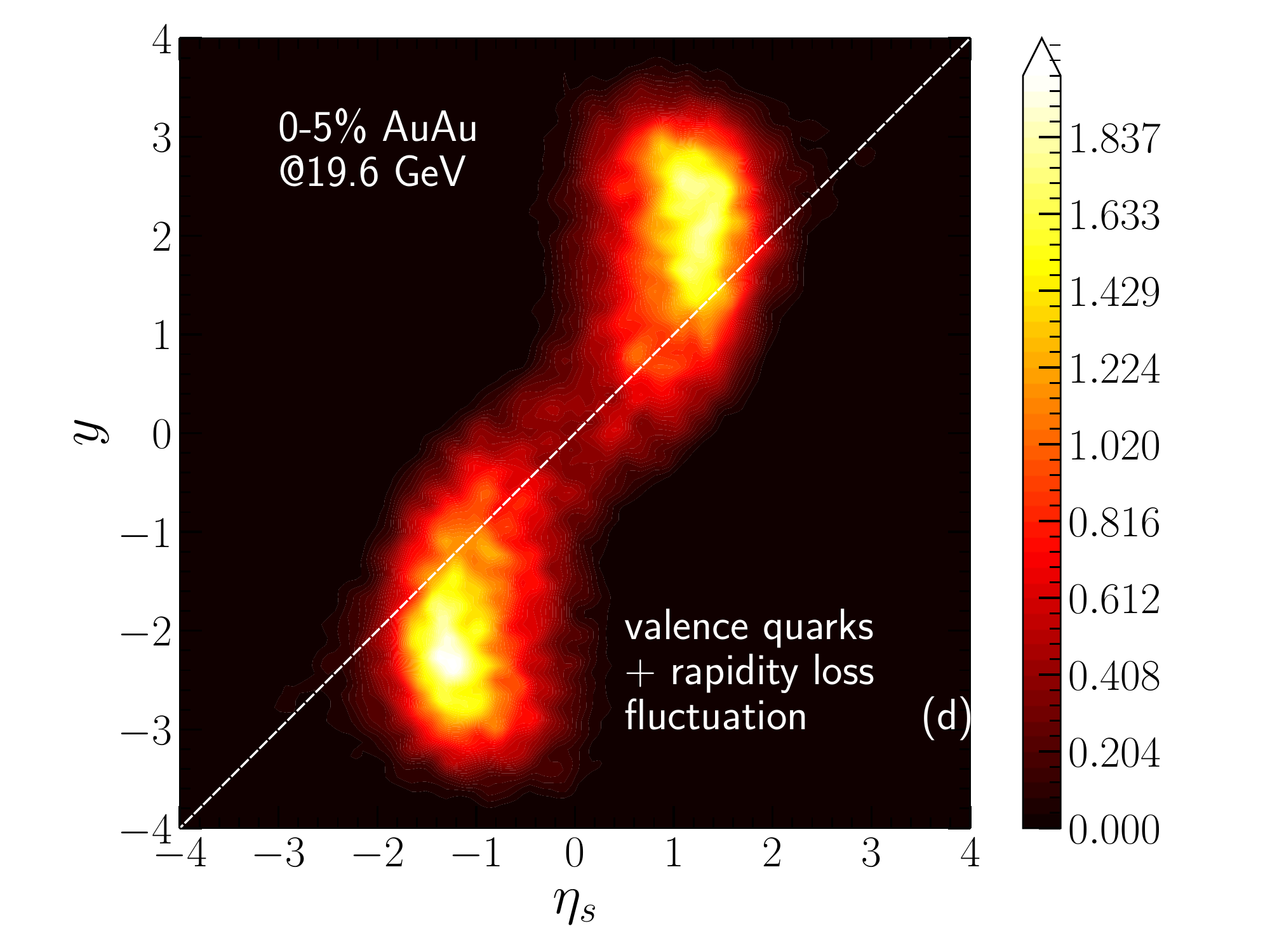}
\end{tabular}
\caption{Panels (a) and (b): The event-averaged rapidity and space-time rapidity distributions for the decelerated baryons in central Au+Au collisions at 4 collision energies. Simulations are done with the valence quark + rapidity loss fluctuation model \cite{Shen:2017bsr}. Panels (c) and (d): The correlation between the baryons' rapidity and space-time rapidity from the valence quark and valence quark + rapidity loss fluctuation model, respectively.}
\label{fig3}
\end{center}
\end{figure}
Fig.~\ref{fig3} shows the decelerated baryon rapidity and space-time rapidity distributions in central Au+Au collisions at 4 collision energies. There are two ways to slow down the baryons in our models. Firstly, the longitudinal momentum fluctuations from sampling valence quarks allows baryons to have some small incoming velocities. Secondly, the additional rapidity loss fluctuation allows a longer deceleration time to decelerate baryons to small rapidity. These two ways of stopping baryons lead to different space-momentum correlations between the baryons' rapidity and space-time rapidity, as illustrated in Fig.~\ref{fig3}c and Fig.~\ref{fig3}d. The fluctuations in the valence quark model lead to a smearing of the incoming particle rapidity. A good linear correlation between $y$ and $\eta_s$ is observed in Fig.~\ref{fig3}c. The additional rapidity loss fluctuation allows for a longer string deceleration time than $\Delta \tau = 0.5$ fm. The longer deceleration time diffuses some large rapidity quarks to lower velocities, yet they reach larger space-time rapidities, changing the shape of the correlation.

\begin{figure}[h!]
\begin{center}
 \includegraphics[width=0.6\linewidth]{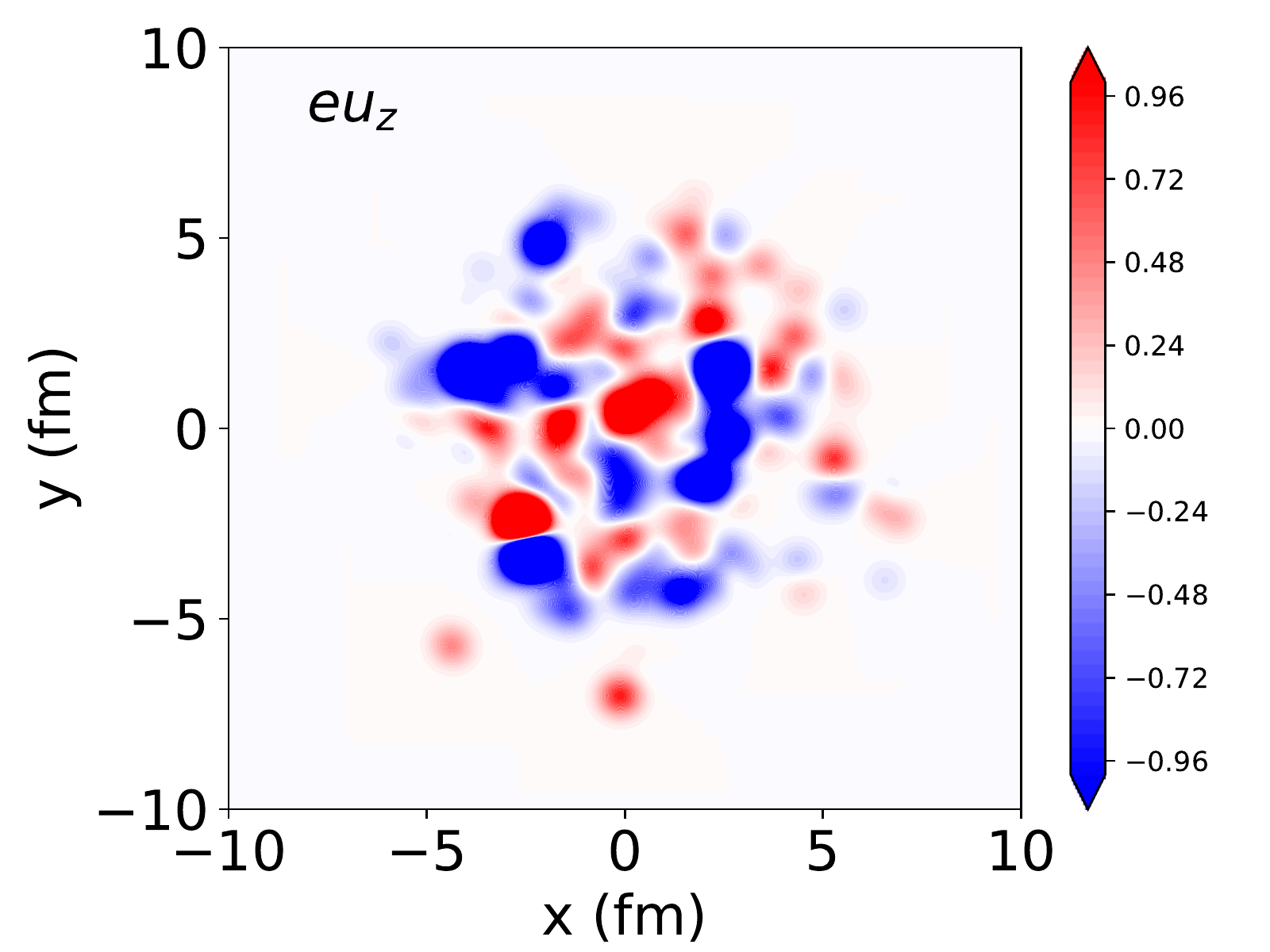}
\caption{The fluctuation of the initial energy flow velocity along the longitudinal direction in one 0-5\% Au+Au collision at 19.6 GeV. }
\label{fig4}
\end{center}
\end{figure}
Decelerated strings are treated as source terms to feed into hydrodynamic simulations \cite{Shen:2017bsr}. As shown in Figs.~\ref{fig3}, the string's rapidity does not coincide with its space-time rapidity at its thermalization. A non-zero longitudinal flow is expected from this model. Fig.~\ref{fig4} shows the longitudinal flow of the energy density at the mid-rapidity region at hydrodynamization. The scale of longitudinal flow fluctuation in the transverse plane is determined by the size of Gaussian width used for the transverse energy density. Similar longitudinal flow fluctuations exist on a much smaller scale in the IP-Glasma model at higher collision energy \cite{McDonald:2017eml}. 
The non-zero longitudinal flow fluctuations in the transverse plane generate a non-zero vorticity for the collision system \cite{Deng:2016gyh,Pang:2016igs}.

\begin{figure}[h!]
\begin{center}
\begin{tabular}{cc}
 \includegraphics[width=0.48\linewidth]{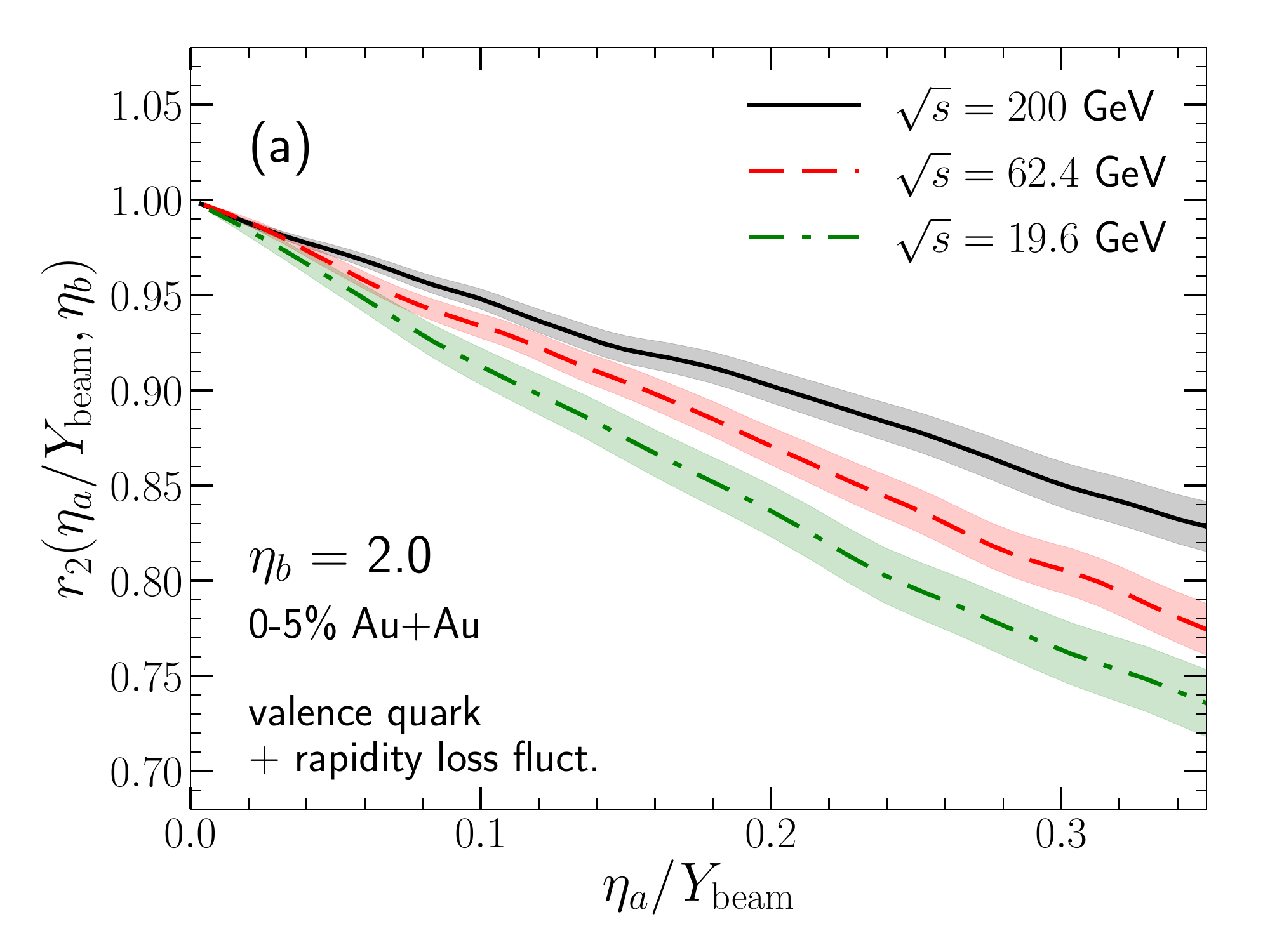} &
 \includegraphics[width=0.48\linewidth]{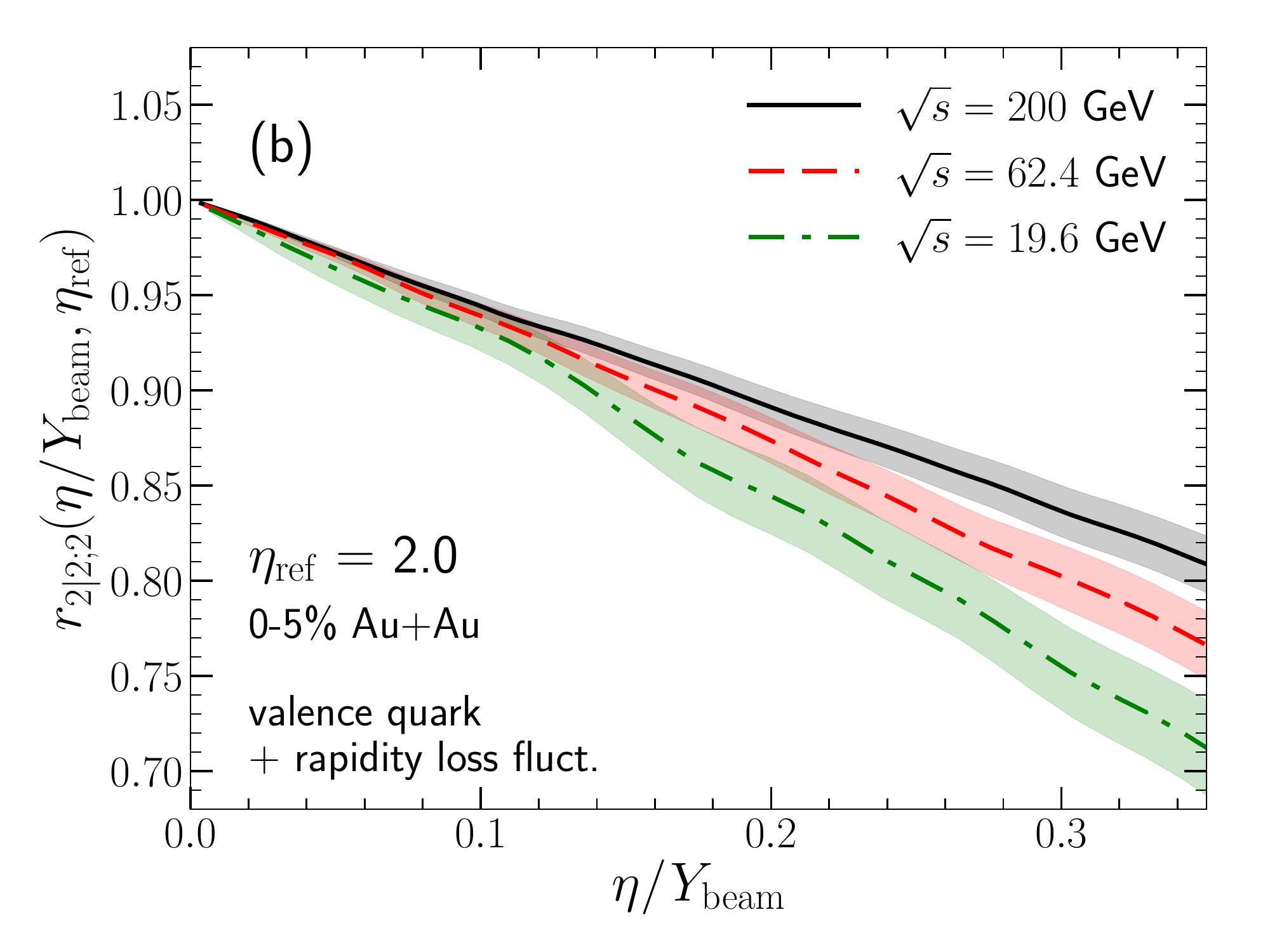}
\end{tabular}
\caption{The longitudinal event-plane decorrelation ratios $r_2$ and $r_{2\vert2;2}$ for the second order spatial eccentricity in 0-5\% Au+Au collisions at three collision energies from the valence quark + rapidity loss fluctuation model. The decorrelation ratios are scaled by the beam rapidity $Y_\mathrm{beam} = \mathrm{arccosh}(\sqrt{s_\mathrm{NN}}/(2m_p))$.}
\label{fig5}
\end{center}
\end{figure}
Longitudinal fluctuations decorrelate the spatial eccentricities $\mathcal{E}_n$ in different rapidity regions. This can be quantified by the following correlation functions \cite{Jia:2015jga,Jia:2017kdq},
\begin{equation}
r_n (\eta_a, \eta_b) = \frac{\langle \Re\{ \mathcal{E}_n(-\eta_a) \cdot \mathcal{E}^*_n(\eta_b) \} \rangle_\mathrm{ev} }{\langle \Re\{ \mathcal{E}_n(\eta_a) \cdot \mathcal{E}^*_n(\eta_b) \} \rangle_\mathrm{ev}}
\end{equation}
and
\begin{equation}
r_{n \vert n;2}(\eta, \eta_\mathrm{ref}) = \frac{\langle \Re \{\mathcal{E}_n(-\eta_\mathrm{ref}) \cdot \mathcal{E}^*_n(\eta)  \mathcal{E}_n(-\eta) \cdot \mathcal{E}^*_n(\eta_\mathrm{ref}) \} \rangle_\mathrm{ev}}{\langle \Re \{\mathcal{E}_n(-\eta_\mathrm{ref}) \cdot \mathcal{E}^*_n(-\eta)  \mathcal{E}_n(\eta) \cdot \mathcal{E}^*_n(\eta_\mathrm{ref}) \} \rangle_\mathrm{ev}}.
\end{equation}
The correlation function $r_{n\vert n;2}(\eta, \eta_\mathrm{ref})$ is mostly sensitive to the event-plane angle decorrelation in different rapidity regions \cite{Jia:2017kdq}, while the ratio $r_n$ is also sensitive to the $\mathcal{E}_n$ magnitude fluctuations. 
In Fig.~\ref{fig5} we present the $r_2$ and $r_{2\vert2;2}$ correlation functions at different collision energies, scaled by their corresponding beam rapidity. We find a larger event-plane decorrelation at lower collision energies even after scaling with the beam rapidity. The value of $r_2$ and $r_{2\vert2;2}$ are very close to each other, which indicates that the decorrelation is mostly from the angular fluctuations in the longitudinal direction.

\section{Conclusion}

In this proceeding we discussed some details about the dynamical initialization framework proposed in Ref.~\cite{Shen:2017bsr}. This model can be applied to simulate the dynamics of relativistic heavy-ion collisions at RHIC BES energies. It is a fully 3D initial condition model which interweaves with relativistic hydrodynamics locally on a collision-by-collision basis. Such a theoretical framework can help us quantify the importance of the pre-equilibrium stage at low collision energies. The deceleration dynamics was employed to generate realistic spatial-momentum correlations for baryons. We introduced two sources of longitudinal fluctuations from the incoming valence quark fluctuations and collision-by-collision rapidity loss fluctuations. The amount of fluctuations can be constrained by comparing with experimental measurements in the RHIC BES program. 
In the future we will integrate this theoretical framework with advanced statistical tools such as the Bayesian analysis \cite{Bernhard:2016tnd} and deep learning techniques \cite{Pang:2016vdc}. A global fit to a variety of RHIC BES measurements can constrain initial state fluctuations and extract the transport properties of the Quark-Gluon Plasma in a baryon rich environment. This is a crucial step towards unraveling the phase structure of QCD and determining the existence and position of the critical point.

\section*{Acknowledgments}
We thank Volker Koch for useful discussions. CS and BPS are supported under DOE Contract No. DE-SC0012704. This research used resources of the National Energy Research Scientific Computing Center, which is supported by the Office of Science of the U.S. Department of Energy under Contract No. DE-AC02-05CH11231. BPS acknowledges a DOE Office of Science Early Career Award. CS gratefully acknowledges a Goldhaber Distinguished Fellowship from Brookhaven Science Associates.
This work is supported in part by the U.S. Department of Energy, Office of Science, Office of Nuclear Physics, within the framework of the Beam Energy Scan Theory (BEST) Topical Collaboration.

\end{document}